\documentclass[aps,pra,twocolumn,reprint,showpacs,floatfix,longbibliography]{revtex4}
\usepackage[utf8]{inputenc} 
\usepackage{graphicx}
\usepackage{amsmath}
\usepackage{amsfonts}
\usepackage{mathtools}
\usepackage{float}
\usepackage{tabularx, booktabs}
\usepackage{color}
\usepackage{braket}

\begin{document}
\title{Three-body interaction near a narrow two-body zero crossing}
\author{A. Pricoupenko and D. S. Petrov}
\affiliation{LPTMS, CNRS, Univ.~Paris-Sud, Universit\'{e} Paris-Saclay, F-91405 Orsay, France}

\date{\today}

\begin{abstract}

We calculate the effective three-body force for bosons interacting with each other by a two-body potential tuned to a narrow zero crossing in any dimension. We use the standard two-channel model parametrized by the background atom-atom interaction strength, the amplitude of the open-channel to closed-channel coupling, and the atom-dimer interaction strength. The three-body force originates from the atom-dimer interaction, but it can be dramatically enhanced for narrow crossings, i.e., for small atom-dimer conversion amplitudes. This effect can be used to stabilize quasi-two-dimensional dipolar atoms and molecules.

\end{abstract}

 \maketitle

\section{Introduction}

In recent years, dilute weakly interacting bosons with intentionally weakened mean-field interactions have become one of the main attractions in the field of quantum gases. The weakness of the mean-field interaction in such systems makes higher-order terms relatively more important leading to dramatic effects. A prominent example is the observation of dilute quantum droplets in dipolar atoms \cite{Kadau2016,Schmitt2016,Ferrier2016,Chomaz2016} and in nondipolar mixtures \cite{Cabrera2017,Cheiney2018,Semeghini2018}. Two-body interactions of different kinds (contact and dipole-dipole in the dipolar case and interspecies and intraspecies in the mixture case) are tuned to compete with each other such that the resulting weak overall attraction gets compensated by a higher-order Lee-Huang-Yang (LHY) term \cite{PetrovLHY2015,Wachtler2016,Bisset2016}. An impressive experimental progress has been made in the dipolar case on pursuing supersolidity through the formation of coherent arrays of quantum droplets \cite{Tanzi2019,Chomaz2019,Bottcher2019,Tanzi2019Modes,Guo2019}. 

All these achievements correspond to essentially three-dimensional setups well described by the Gross-Pitaevskii energy density functional with an additional local LHY term $\propto n^{5/2}$, where $n$ is the density. However, there are various reasons to consider other configurations where the $n^{5/2}$ term is absent or too weak (low-dimensional geometries, single-component contact-interacting atoms, etc.) In these cases, an effective three-body interaction, associated with a $n^3$ term in the energy density, can become dominant if the leading-order two-body forces are suppressed. In particular, three-body forces have been considered in the context of droplet formation in three dimensions \cite{Bulgac2002,PetrovThreeBody2014,Xi2016,Bisset2015} and as a means for stabilizing supersolid phases of quasi-two-dimensional dipolar atoms or molecules \cite{Lu2015}. Quite a few recent theory papers have discussed one-dimensional three-body-interacting systems, exploring the kinematic equivalence of the three-body scattering in one dimension and the two-body scattering in two dimensions (see, for example, Refs.~\cite{PricoupenkoJr2018,Sekino2018,Drut2018,Nishida2018,Pricoupenko2018,Guijarro2018,Pricoupenko2019,Daza2018,McKenney2019,Pastukhov2019,Valiente2019,ValientePastukhov2019}).

In this paper we analyze a simple mechanism for the emergence of an effective three-body interaction. Namely, we consider bosons interacting with each other by a potential tuned to a zero crossing near a narrow Feshbach resonance, where the conversion amplitude from atoms to closed-channel dimers is small and where the two-body scattering amplitude is characterized by a large effective range $R_{\rm e}$. The effective three-body force appears in this model when one takes into account the interaction between atoms and closed-channel dimers, characterized by the coupling strength $g_{12}$. We find that the three-body coupling constant $g_3$ in $D$ dimensions is proportional to $R_{\rm e}^Dg_{12}$ and can thus be enhanced near narrow two-body zero crossings. 

The paper is organized as follows. In Sec.~\ref{MF} we introduce the two-channel model and perform its mean-field analysis. In the dilute limit the density of closed-channel dimers in the system scales as $R_{\rm e}^Dn^2\ll n$ and the effective three-body interaction emerges simply as the atom-dimer mean-field interaction energy $\propto R_{\rm e}^D g_{12}n^3$. We show that this simple mechanism, applied to two-dimensional dipoles, generates conditions for observing supersolid phases predicted in Ref.~\cite{Lu2015}. 

In Secs.~\ref{TwoBody} and \ref{ThreeBody} we turn to the few-body perspective and perform a detailed nonperturbative analysis of the two-body (Sec.~\ref{TwoBody}) and three-body (Sec.~\ref{ThreeBody}) problems with zero-range potentials. In particular, the three-body scattering length near a narrow two-body zero crossing is found for an arbitrary atom-dimer interaction strength in any dimension.

\section{Mean-field analysis\label{MF}}

We start with the two-channel model described by the Hamiltonian \cite{Radzihovsky2008}
\begin{align}
\hat{H}=\int_{\bf r} \Big\{-\hat{\psi}_1^\dagger({\bf r}) \frac{\nabla^2}{2}\hat{\psi}_1({\bf r})+\hat{\psi}_2^\dagger({\bf r}) \left(-\frac{\nabla^2}{4}+\nu_0\right) \hat{\psi}_2({\bf r})\nonumber\\
-\frac{\alpha}{2}[\hat{\psi}_1^\dagger({\bf r}) \hat{\psi}_1^\dagger({\bf r})\hat{\psi}_2({\bf r})+{\rm h.c.}]+\sum_{\sigma\sigma'}\frac{g_{\sigma\sigma'}}{2}\hat{n}_\sigma({\bf r}) \hat{n}_{\sigma'}({\bf r})\Big\},\label{Ham}
\end{align}
where $\hat{\psi}_1$ and $\hat{\psi}_2$ are, respectively, the annihilation operators of atoms and dimers, $\hat{n}_\sigma$ are the corresponding density operators, $\nu_0$ is the detuning parameter, $g_{\sigma\sigma'}$ are interaction constants, $\alpha$ is the atom-dimer conversion amplitude (without loss of generality assumed real and positive), and we have set $\hbar$ and atom mass equal to 1. Hereafter, $\int_{\bf r}$ denotes $\int d^Dr$.

In the mean-field description of (\ref{Ham}) we assume pure atomic and molecular condensates $\hat{\psi}_\sigma=\sqrt{n_\sigma}$ with the same phase (which corresponds to the energy minimum for $\alpha>0$) \cite{Radzihovsky2008}. We arrive at the energy density
\begin{equation}\label{EnDens}
E/L^D=\nu_0 n_2-\alpha n_1 \sqrt{n_2}+\sum_{\sigma\sigma'}g_{\sigma\sigma'}{n}_\sigma {n}_{\sigma'}/2,
\end{equation}
which we minimize with respect to $n_2$ (or $n_1$) keeping the total density $n=n_1+2n_2$ constant. For positive $\nu_0$ and small $n$ the dimer population behaves quadratically in $n$ 
\begin{equation}\label{n2expansion}
n_2=\left(\frac{\alpha n}{2\nu_0}\right)^2\left(1+\frac{4g_{11}\nu_0-2g_{12}\nu_0-3\alpha^2}{\nu_0^2}n\right)+O(n^4)
\end{equation}
and the energy density reads
\begin{equation}\label{En}
\frac{E}{L^D}=\left(\frac{g_{11}}{2}-\frac{\alpha^2}{4\nu_0}\right)\left(n^2-\frac{\alpha^2}{\nu_0^2}n^3\right)+\frac{g_{12}\alpha^2}{4\nu_0^2}n^3+O(n^4).
\end{equation} 

The two-body zero crossing occurs at the detuning $\nu_0=\alpha^2/2g_{11}$ where the first term in the right-hand side of Eq.~(\ref{En}) vanishes. One can then see that the residual three-body energy shift originates from the direct mean-field interaction of atoms with dimers. It equals $g_{12} n_1 n_2 \approx g_3n^3/3!$ with 
\begin{equation}\label{g3}
g_3=6g_{12}g_{11}^2/\alpha^2=3g_{12}R_{\rm e}^D.
\end{equation} 
The effective volume $R_{\rm e}^D=2g_{11}^2/\alpha^2$ introduced in Eq.~(\ref{g3}) characterizes the closed-channel population. Indeed, the density of dimers can be written as 
\begin{equation}\label{n2}
n_2\approx R_{\rm e}^Dn^2/2
\end{equation}
meaning that each pair of atoms is found in the closed-channel dimer state with probability $(R_{\rm e}/L)^D$. 

If $g_{\sigma\sigma'}$ are of the same order of magnitude $\sim g$, the expansion (\ref{En}) is in powers of $R_{\rm e}^D n$, which we assume small. Then, at the zero crossing the three-body term gives the leading contribution to the energy density  $\sim gn^2 (R_{\rm e}^{D}n)^1$ and we neglect subleading terms such as, for instance, the dimer-dimer interaction $\sim g_{22}\alpha^4n^4/\nu_0^4\sim gn^2(R_{\rm e}^{D}n)^2$. On the other hand, it may be interesting to keep a small but finite effective two-body interaction $g_{\rm eff}=g_{11}-\alpha^2/2\nu_0 \sim g(R_{\rm e}^{D}n)\ll g$, so that it can compete with the three-body term. It is also useful to note that the effective two-body interaction depends on the collisional momentum as $g_{\rm eff}(k)=g_{\rm eff}(0)-R_{\rm e}^Dk^2$ (see \cite{Shotan2014} and Sec.~\ref{TwoBody}). However, if $k\ll \sqrt{gn}$, the corresponding effective-range correction gives a contribution to (\ref{En}) much smaller than $gn^2 (R_{\rm e}^{D}n)^1$. We thus conclude that on this level of expansion we reduce (\ref{Ham}) to the model of scalar bosons with local effective two-body and three-body interactions.

\subsection{Application to two-dimensional dipoles}

Having in mind supersolid phases, which require a three-body repulsive force \cite{Lu2015}, let us perform the same mean-field analysis in the case of two-dimensional dipoles oriented perpendicular to the plane. Instead of pointlike interactions characterized by the momentum-independent constants $g_{\sigma\sigma'}$ we now assume momentum-dependent pseudopotentials \cite{Baranov2011,Lu2015}
\begin{equation}\label{Vtilde}
\tilde{V}_{\sigma\sigma'}(|{\bf k}-{\bf k}'|)=g_{\sigma\sigma'}-2\pi d_\sigma d_{\sigma'}|{\bf k}-{\bf k}'|,
\end{equation}
where ${\bf k}$ and ${\bf k}'$ are the incoming and outgoing relative momenta and $d_1$ and $d_2$ are dipole moments of atoms and dimers, respectively. The pseudopotential (\ref{Vtilde}) is an effective potential valid only for the leading-order mean-field analysis at low momenta. Its coordinate representation
\begin{equation}\label{V}
V_{\sigma\sigma'}({\bf r}-{\bf r}')=\int \frac{d^2q}{(2\pi)^2}\tilde{V}_{\sigma\sigma'}(q)e^{i{\bf q}({\bf r}-{\bf r}')}
\end{equation}
has the long-distance asymptote $d_\sigma d_{\sigma'}/r^3$ with the characteristic range $r^{*}_{\sigma\sigma'}=2\mu_{\sigma\sigma'}d_\sigma d_{\sigma'}$, where $\mu_{11}=1/2$ and $\mu_{12}=2/3$ are the atom-atom and atom-dimer reduced masses, respectively.

Obviously, for homogeneous condensates the momentum-dependent part of (\ref{Vtilde}) plays no role and our previous analysis holds. Namely, we arrive at the energy density $E/L^2=g_{\rm eff}n^2/2+g_3 n^3/6$, where $g_{\rm eff}=g_{11}-\alpha^2/2\nu_0$ is tuned to be small and $g_3$ is given by Eq.~(\ref{g3}). Let us now assume that the atomic and dimer condensates are spatially modulated with a characteristic momentum $k$ (in the supersolid phase the modulation is periodic). Then, the most important new terms in Eqs.~(\ref{EnDens}) and (\ref{En}) are the kinetic energy of the atomic component $\sim n k^2$ and the momentum-dependent part of the atom-atom interaction $\sim -r_{11}^{*}k n^2$. Minimizing their sum with respect to $k$ gives a contribution $\epsilon_{\rm mod}\sim -r_{11}^{*2}n^3$ to the energy density and the optimal modulation momentum $k_{\rm min}\sim r_{11}^{*}n$ \cite{Lu2015}. One can check that other momentum-dependent terms are subleading. For instance, the kinetic energy of dimers $\sim n_2 k^2$ and the momentum-dependent atom-dimer interaction $\sim r_{12}^{*}kn n_2$ carry an additional factor $R_{\rm e}^2n\ll 1$. It is important to mention that the density of dimers satisfies Eq.~(\ref{n2}) locally, i.e., $n_2({\bf r})\approx R_{\rm e}^2n^2({\bf r})/2$. Deviations from this relation, which follows from minimizing the first two terms in the right-hand side of Eq.~(\ref{EnDens}), are energetically too costly. A change of $n_2$ by, say, a factor of two compared to the optimal value would cost $\sim g_{11}n^2\gg gn^2(R_{\rm e}^2n)$ in the energy density.

This analysis leads us to the model of two-dimensional dipoles characterized by an effective two-body pseudopotential $\tilde{V}(k)=g_{\rm eff}-2\pi d_1^2k$ and local three-body term $g_3\delta({\bf r}_1-{\bf r}_2)\delta({\bf r}_2-{\bf r}_3)$. The mean-field phase diagram of this model has been worked out in Ref.~\cite{Lu2015}. It has been shown that the stability of the system with respect to collapse is ensured by the repulsive three-body interaction term compensating the effectively attractive $\epsilon_{\rm mod}$, which also scales as $n^3$. The supersolid stripe, honeycomb and triangular phases are predicted when these two terms are comparable and $g_{\rm eff}<0$. To give a concrete example, the four-critical point where the three supersolid phases meet with one another and with the uniform phase (this is also the point where the roton minimum touches zero) is characterized by $g_{12}R_{\rm e}^2=2(\pi r_{11}^{*})^2$ and $nR_{\rm e}^2=|g_{\rm eff}|/g_{12}$.

\subsection{Inelastic losses}

Collisions of atoms with closed-channel dimers can lead to the relaxation to more deeply bound molecular states. The rate of this process in a unit volume is given by $\alpha_{\rm r}n_1n_2$, where $\alpha_{\rm r}$ is the relaxation rate constant. In our model this corresponds to the atom loss rate $\dot{n}=-(3/2)\alpha_{\rm r} R_{\rm e}^D n^3$, and we see that this effective three-body loss gets enhanced with increasing $R_{\rm e}$ in the same manner as the elastic three-body interaction (\ref{g3}). In fact, the atom-dimer relaxation can be mathematically modeled by allowing $g_{12}$ to be complex. Shotan and co-workers \cite{Shotan2014} have measured the three-body loss rate constant near a two-body zero crossing in three dimensions. They argue that this quantity is proportional to $R_{\rm e}^4$. Here we claim a slightly different scaling ($\propto R_{\rm e}^3$), valid when $R_{\rm e}$ is much larger than the van der Waals range.

For Feshbach molecules of the size of the van der Waals length $\alpha_{\rm r}$ is typically of the same order of magnitude as $g_{12}$. The lifetime of the sample is thus comparable to the timescale associated with the elastic three-body energy shift. There are, however, ways of overcoming this problem. For dipoles oriented perpendicular to the plane in the quasi-two-dimensional geometry inelastic processes are suppressed by the predominantly repulsive dipolar tail. For instance, for Dy the atom-dimer dipolar length $r_{12}^{*}$ can reach about 50~nm depending on the magnetic moment of the closed-channel dimer. The confinement of frequency $\omega=2\pi\times$100~kHz for this system gives the oscillator length $\sqrt{\hbar/2\mu_{12}\omega}\approx 21$~nm. Under these conditions one expects a noticeable reduction of the relaxation rate \cite{Quemener2010,Micheli2010,Frisch2015}. This mechanism may work also for dipolar molecules where larger values of $r_{12}^{*}$ can be reached. 

A different approach to this problem is to consider closed-channel dimers which are weakly-bound and have a halo character, i.e., well extended beyond the support of the potential. A specific way of generating three-body interactions in this manner has been proposed by one of us in Ref.~\cite{PetrovThreeBody2014}; two atoms in state 1 collide and both go to another internal state $1'$ where they form an extended molecular state. The effective three-body force is then due to a repulsive mean-field interaction between atoms $1'$ and a third atom in state 1. In this case, the relaxation is slow since the dimer is not ``preformed''. 
     
\section{Regularized model and two-body problem\label{TwoBody}}    

We now go back to the model (\ref{Ham}), try to analyze it from the few-body viewpoint, and characterize the three-body interaction beyond the mean-field result (\ref{g3}) (also trying to determine its validity regime). Clearly, at some point the strength of the background atom-atom interaction becomes a relevant parameter (not just the ratio $g_{11}/\alpha$). One also observes that the pointlike interaction and conversion terms in Eq.~(\ref{Ham}) lead to divergences and have to be regularized in dimensions $D>1$, which necessitates an additional parameter (a short-range or high-momentum cutoff).

In order to regularize the model (\ref{Ham}) we use the delta-shell pseudopotential representation \cite{Stock2005,Kanjilal2006} with a finite range $r_0$. Namely, we rewrite Eq.~(\ref{Ham}) as
\begin{align}
\hat{H}=\int_{\bf r} -\hat{\psi}_1^\dagger({\bf r}) \frac{\nabla^2}{2}\hat{\psi}_1({\bf r})+\hat{\psi}_2^\dagger({\bf r}) \left(-\frac{\nabla^2}{4}+\nu_0\right) \hat{\psi}_2({\bf r})\nonumber\\
+\sum_{\sigma\sigma'}\frac{g_{\sigma\sigma'}}{2}\int_{\bf r}\int_{\bf y} \tilde\delta_{r_0}({\bf y})\hat{n}_\sigma({\bf r}+{\bf y}/2)\hat{n}_{\sigma'}({\bf r}-{\bf y}/2)\nonumber\\
-\frac{\alpha}{2} \int_{\bf r} \int_{\bf y} \tilde\delta_{r_0}({\bf y}) [\hat{\psi}^\dagger_1({\bf r}+{\bf y}/2)\hat{\psi}^\dagger_{1}({\bf r}-{\bf y}/2)\hat{\psi}_2({\bf r})+{\rm h.c.}],\label{Hamr0}
\end{align}
where $\tilde\delta_{r_0}({\bf y}) = \delta(|{\bf y}|-r_0)/S_D(r_0)$ is the normalized delta shell with $S_1(r_0)=2$, $S_2(r_0)=2\pi r_0$, and $S_3(r_0)=4\pi r_0^2$. The range $r_0$ should be understood as the smallest lengthscale in our problem. It does not enter in the final formulas and it is just a convenient way to regularize the problem without using zero-range pseudopotentials, which have different forms in different dimensions. In the one-dimensional case $r_0$ can be set to zero from the very beginning, but we keep it finite in order to use the same formalism for the cases with different $D$. Note also that we do not intend to consider effects of scattering with angular momenta $l\neq 0$. This is to say that, as $r_0$ is decreased, the coupling constants $g_{\sigma\sigma'}$ and $\alpha$ are tuned to reproduce desired (physical) $R_{\rm e}$ and $a_{\sigma\sigma'}$ only for the $s$-wave channel. Then, in the limit $r_0\rightarrow 0$, the terms  $g_{\sigma\sigma'}\tilde\delta_{r_0}({\bf y})$ and $\alpha\tilde\delta_{r_0}({\bf y})$ are too weak to induce any scattering for $l>0$.

A stationary two-body state with zero center-of-mass momentum and $l=0$ in the two-channel models (\ref{Ham}) or (\ref{Hamr0}) is represented by
\begin{equation}\label{TwoBodyState}
\int_{\bf c}\int_{\bf y} \Psi(y) \hat{\psi}^\dagger_1({\bf c}+{\bf y}/2)\hat{\psi}^\dagger_1({\bf c}-{\bf y}/2)\ket{0}+\int_{\bf c}\phi \hat{\psi}^\dagger_2({\bf c})\ket{0},
\end{equation}
where $\ket{0}$ is the vacuum state. Acting on (\ref{TwoBodyState}) by the operator $\hat{H}-E$, and requiring that the result vanish, we get the coupled Schr\"odinger equations at energy $E$, 
\begin{align}
&[-\nabla^2_{\bf y}-E+g_{11}\tilde\delta_{r_0}(y)]\Psi(y)=\alpha\tilde\delta_{r_0}(y)\phi/2,\label{TwoBodyPsi}\\
&(\nu_0-E)\phi=\alpha\Psi(r_0),\label{TwoBodyphi}
\end{align}
which, upon eliminating the closed-channel amplitude $\phi$, become
\begin{equation}\label{TwoBodyEff}
[-\nabla^2_{\bf y}-E+g_{\rm eff}(E)\tilde\delta_{r_0}(y)]\Psi(y)=0
\end{equation}
with 
\begin{equation}\label{geff}
g_{\rm eff}(E)=g_{11}+\frac{1}{2}\frac{\alpha^2}{E-\nu_0}.
\end{equation}
The zero crossing condition at zero energy thus reads
\begin{equation}\label{ZeroCrossCond}
\nu_0=\alpha^2/2g_{11}.
\end{equation} 
We also introduce the effective range by the formula
\begin{equation}\label{EffRange}
R_{\rm e}^D=\alpha^2/2\nu_0^2>0,
\end{equation}
which characterizes the small-$E$ asymptote $g_{\rm eff}(E)=g_{\rm eff}(0)-R_{\rm e}^DE+O(E^2)$ (cf. \cite{Shotan2014}). At the crossing Eq.~(\ref{EffRange}) is consistent with our earlier definition of $R_{\rm e}$ introduced in Eq.~(\ref{g3}). As we have mentioned, $R_{\rm e}^D$ is also related to the closed-channel occupation. Indeed, from the normalization integral of Eq.~(\ref{TwoBodyState}) one finds that the closed-channel to open-channel probability ratio equals $|\phi|^2/\int_{\bf y}2|\Psi(y)|^2=|\phi|^2/(2L^D|\Psi(r_0)|^2)$ where we have used the fact that at the crossing $\Psi(y)=\Psi(r_0)$. On the other hand, from Eq.~(\ref{TwoBodyphi}) one obtains $|\phi|^2=2R_{\rm e}^D|\Psi(r_0)|^2$ for $|E|\ll |\nu_0|$, which gives the result claimed in Sec.~\ref{MF}. Namely, the probability for two atoms to be in the closed-channel dimer state equals $(R_{\rm e}/L)^D$.

Eventually, we will need to express our results in terms of the scattering lengths $a_{\sigma\sigma'}$ and the effective range $R_{\rm e}$ rather than in terms of the bare $r_0$-dependent quantities $g_{\sigma\sigma'}$, $\alpha$, and $\nu_0$. Relations between $g_{\sigma\sigma'}$ and $a_{\sigma\sigma'}$ are obtained by solving the scattering problem at zero collision energy and by looking at the long-distance asymptote of the two-body wave function. Namely, the zero-energy Schr\"odinger equation reads
\begin{equation}\label{TwoBodysigma}
[-\nabla^2_{\bf y}+2\mu_{\sigma\sigma'}g_{\sigma\sigma'}\tilde\delta_{r_0}(y)]\Psi(y)=0.
\end{equation}
In one dimension the (unnormalized) solution is 
\begin{equation}\label{Psi1D}
\Psi(y)=\left\{\begin{aligned} 1,\;|y|<r_0\\1+\mu_{\sigma\sigma'}g_{\sigma\sigma'}(|y|-r_0),\;|y|>r_0,\end{aligned}\right.
\end{equation}
from which we see that 
\begin{equation}\label{g1D}
a_{\sigma\sigma'}=r_0-1/\mu_{\sigma\sigma'}g_{\sigma\sigma'}.
\end{equation} 
In the limit $r_0\rightarrow 0$ we recover the usual relation $g_{\sigma\sigma'}=-1/\mu_{\sigma\sigma'}a_{\sigma\sigma'}$.
In two dimensions the solution of Eq.~(\ref{TwoBodysigma}) reads
\begin{equation}\label{Psi2D}
\Psi({\bf y})=\left\{\begin{aligned} 1,\;|{\bf y}|<r_0\\1+\mu_{\sigma\sigma'}g_{\sigma\sigma'}\ln(|{\bf y}|/r_0)/\pi,\;|{\bf y}|>r_0\end{aligned}\right.
\end{equation}
and one has 
\begin{equation}\label{g2D}
\mu_{\sigma\sigma'}g_{\sigma\sigma'}=\pi/\ln(r_0/a_{\sigma\sigma'}).
\end{equation}
In three dimensions
\begin{equation}\label{Psi3D}
\Psi({\bf y})=\left\{\begin{aligned} 1,\;|{\bf y}|<r_0\\1-\mu_{\sigma\sigma'}g_{\sigma\sigma'}/2\pi|{\bf y}|+\mu_{\sigma\sigma'}g_{\sigma\sigma'}/2\pi r_0,\;|{\bf y}|>r_0,\end{aligned}\right.
\end{equation}
from which we obtain 
\begin{equation}\label{g3D}
1/a_{\sigma\sigma'}=2\pi/\mu_{\sigma\sigma'}g_{\sigma\sigma'}+1/r_0.
\end{equation}

We now analyze conditions for having two-body bound states at the two-body zero crossing, in particular, having in mind the three-body recombination to these states when considering the three-body problem. We just note that solutions of Eq.~(\ref{TwoBodyEff}) at distances $|y|\ll 1/\sqrt{|E|}$ in different dimensions are given, respectively, by Eqs.~(\ref{Psi1D}), (\ref{Psi2D}), and (\ref{Psi3D}) with $\sigma=\sigma'=1$ and with $g_{11}$ substituted by $g_{\rm eff}(E)$. We then match these asymptotes with the decaying solutions $\Psi^{(D=1)}(y)\propto \exp (\kappa |y|)$, $\Psi^{(D=2)}(y)\propto K_0 (\kappa |{\bf y}|)$, and $\Psi^{(D=3)}(y)\propto \exp (-\kappa |{\bf y}|)/|{\bf y}|$, where $\kappa=\sqrt{-E}$. This matching procedure gives the following equations for the determination of $\kappa$ ($\gamma\approx 0.577$ is the Euler constant):
\begin{align}
&(\kappa R_{\rm e})^2(a_{11}/R_{\rm e})-\kappa R_{\rm e}=2,& D=1,\label{kappa1D}\\
&(\kappa R_{\rm e})^2\ln(\kappa a_{11} e^\gamma/2)=2\pi,& D=2,\label{kappa2D}\\
&(\kappa R_{\rm e})^3-(\kappa R_{\rm e})^2(R_{\rm e}/a_{11})=4\pi,& D=3.\label{kappa3D}
\end{align}
Analyzing these equations we find that in one dimension there is no two-body bound state, if $a_{11}<0$ (or $g_{11}>0$). In higher dimensions we always have a bound state, but it becomes deep in the limit of small positive $a_{11}$ ($E\propto -1/a_{11}^2$). In principle, the case of a weak repulsive background atom-atom interaction can also be realized by a finite-range repulsive potential (in the mean-field spirit of Sec.~\ref{MF}). Then, the dimer states given by Eqs.~(\ref{kappa1D}-\ref{kappa3D}) are spurious, consistent with the fact that the zero-range theory can no longer be used at such high momenta. 

\section{Three-body problem\label{ThreeBody}}

Similar to Eq.~(\ref{TwoBodyState}) a stationary state of three atoms with zero center-of-mass momentum can be written in the form
\begin{widetext}
\begin{equation}\label{ThreeBodyState}
\int_{\bf c}\int_{{\bf x}}\int_{{\bf y}} \Psi({\bf x},{\bf y}) \hat{\psi}^\dagger_1({\bf c}-{\bf x}/2\sqrt{3}-{\bf y}/2)\hat{\psi}^\dagger_1({\bf c}-{\bf x}/2\sqrt{3}+{\bf y}/2)\hat{\psi}^\dagger_1({\bf c}+{\bf x}/\sqrt{3})\ket{0}+\int_{\bf c}\int_{{\bf x}}\phi({\bf x}) \hat{\psi}^\dagger_2({\bf c}-{\bf x}/2\sqrt{3})\hat{\psi}^\dagger_1({\bf c}+{\bf x}/\sqrt{3})\ket{0},
\end{equation}
\end{widetext}
where ${\bf c}$ is the center-of-mass coordinate and the relative Jacobi coordinates are
\begin{align}
{\bf x}&=(2{\bf r}_1-{\bf r}_2-{\bf r}_3)/\sqrt{3},\nonumber\\
{\bf y}&={\bf r}_3-{\bf r}_2.\label{Jacobi}
\end{align} 
Let us introduce operators $\hat{P}_+$ and $\hat{P}_-$ which exchange the first atom with the second and the third, respectively. Acting by these operators on an arbitrary function $F({\bf x},{\bf y})$ results in
\begin{equation}\label{P}
\hat{P}_\pm F({\bf x},{\bf y})=F(-{\bf x}/2\mp\sqrt{3}{\bf y}/2,-\sqrt{3}{\bf x}/2\pm{\bf y}/2).
\end{equation}
The open-channel wave function $\Psi({\bf x},{\bf y})$ is invariant with respect to these permutations.

The coupled Schr\"odinger equations for $\Psi$ and $\phi$ read
\begin{widetext}
\begin{align}
&[-\nabla^2_{\bf x}-\nabla^2_{\bf y}-E + g_{11}(1+\hat{P}_+ + \hat{P}_-)\tilde\delta_{r_0}(y)]\Psi({\bf x},{\bf y})=\alpha(1+\hat{P}_+ + \hat{P}_-)\tilde\delta_{r_0}(y)\phi({\bf x})/2,\label{ThreeBodyPsi}\\
&[-\nabla^2_{\bf x}-\nu_0-E+g_{12}\tilde\delta_{r_0}(\sqrt{3}x/2)]\phi({\bf x})=\alpha\Psi({\bf x},r_0),\label{ThreeBodyphi}
\end{align}
\end{widetext}
where $\Psi({\bf x},r_0)$ in the right-hand side of Eq.~(\ref{ThreeBodyphi}) denotes the projection on the $s$-wave channel in the coordinate ${\bf y}$, i.e., the angular average $\langle\Psi({\bf x},r_0\hat{y})\rangle_{\hat{y}}$. The difference between $\Psi({\bf x},r_0\hat{y})$ and $\Psi({\bf x},r_0)$, which accounts for non-$s$-wave scattering channels, vanishes in the limit $r_0\rightarrow 0$ and we will thus make the replacement $\tilde\delta_{r_0}(y)\Psi({\bf x},{\bf y})\rightarrow \tilde\delta_{r_0}(y)\Psi({\bf x},r_0)$ in Eq.~(\ref{ThreeBodyPsi}). Then, it is convenient (the reason will become clear below) to introduce an auxiliary function $f({\bf x})$ such that
\begin{equation}\label{Psi0throughf}
\Psi({\bf x},r_0)=-f({\bf x})/g_{11}+\alpha\phi({\bf x})/2g_{11}.
\end{equation}
We now eliminate $\Psi$ from Eqs.~(\ref{ThreeBodyPsi}) and (\ref{ThreeBodyphi}) in favor of $f$ and thus derive coupled equations for $f$ and $\phi$. To this end we note that with the use of (\ref{Psi0throughf}) Eq.~(\ref{ThreeBodyPsi}) becomes
\begin{equation}\label{Psif}
(-\nabla^2_{\bf x}-\nabla^2_{\bf y}-E)\Psi({\bf x},{\bf y})=(1+\hat{P}_+ + \hat{P}_-)\tilde\delta_{r_0}(y)f({\bf x}).
\end{equation}
Equation~(\ref{Psif}) can now be solved with respect to $\Psi$ by using the Green function $G_E^{(2D)}$ of the 2$D$-dimensional Helmholtz operator in the left-hand side (see, for example, Ref.~\cite{PetrovLesHouches}). This procedure gives
\begin{widetext}
\begin{equation}\label{Psithroughf}
\Psi({\bf x},r_0)=\Psi_0({\bf x},0)+\int_{{\bf x}'}\left\{G_E^{(2D)}[\sqrt{({\bf x}-{\bf x}')^2+r_0^2}]+\sum_{\pm}G_E^{(2D)}(\sqrt{x^2\pm{\bf x}{\bf x}'+{x'}^2})\right\}f({\bf x}'),
\end{equation}
\end{widetext}
where $\Psi_0({\bf x},{\bf y})$ is any solution of $(-\nabla^2_{\bf x}-\nabla^2_{\bf y}-E)\Psi_0({\bf x},{\bf y})=0$. In Eq.~(\ref{Psithroughf}) we have already taken the limit $r_0\rightarrow 0$, where it exists. With the use of Eq.~(\ref{Psithroughf}) the function $\Psi({\bf x},r_0)$ can now be eliminated from Eqs.~(\ref{Psi0throughf}) and (\ref{ThreeBodyphi}). Here we explicitly write down the resulting coupled equations for $f$ and $\phi$ at the two-body zero crossing ($\nu_0=\alpha^2/2g_{11}$) and at zero energy ($E=0$, $\Psi_0=1$):
\begin{align}
&\hat{L}f({\bf x})+f({\bf x})/g_{11}=\phi({\bf x})/\sqrt{2R_{\rm e}^D}-1,\label{Integr}\\
&[-\nabla^2_{\bf x}+g_{12}\tilde\delta_{r_0}(\sqrt{3}x/2)]\phi({\bf x})=-\sqrt{2/R_{\rm e}^{D}}f({\bf x}),\label{Differ}
\end{align}
where $\hat{L}$ is the integral operator in the right-hand side of Eq.~(\ref{Psithroughf}) with $E=0$. We will use the following forms of the zero-energy Green functions
\begin{align}
&G_0^{(2)}(\rho)=-\ln(\rho/R_{\rm e})/2\pi,\label{G2D}\\
&G_0^{(4)}(\rho)=1/4\pi^2\rho^2,\label{G4D}\\
&G_0^{(6)}(\rho)=1/4\pi^3\rho^4.\label{G6D}
\end{align}
Equations~(\ref{Integr}) and (\ref{Differ}) conserve angular momentum and parity. We will be interested in the case of positive parity (for $D=1$) and zero angular momentum (for $D>1$) so that $f({\bf x})=f(x)$ and $\phi({\bf x})=\phi(x)$. Note also that if $g_{12}=0$, the solution of Eqs.~(\ref{Integr}) and (\ref{Differ}) is $f(x)=0$ and $\phi(x)=\sqrt{2R_{\rm e}^D}$ indicating the absence of two-body and three-body interactions.

The quantity that we want to extract from solving Eqs.~(\ref{Integr}) and (\ref{Differ}) is $\tilde{f}(0)=\int_{{\bf x}}f(x)$, which is proportional to the three-body scattering amplitude. Indeed, at large hyperradii $\rho=\sqrt{x^2+y^2}$ Eq.~(\ref{Psithroughf}) gives $\Psi\approx 1+3\tilde{f}(0)G_0^{(2D)}(\rho)$ or, explicitly,
\begin{equation}\label{PsiAsymp}
\Psi=\left\{ 
\begin{aligned}
&1-3\tilde{f}(0)\ln(\rho/R_{\rm e})/2\pi&\propto \ln (\rho/a_3), &\;\;D=1,\\
&1+3\tilde{f}(0)/4\pi^2\rho^2&\propto 1-S_3/\rho^2, &\;\;D=2,\\
&1+3\tilde{f}(0)/4\pi^3\rho^4&\propto 1-\Upsilon_3/\rho^4, &\;\;D=3,
\end{aligned}
\right.
\end{equation}
where we have introduced the three-body scattering length $a_3$ in one dimension, surface $S_3$ in two dimensions, and hypervolume $\Upsilon_3$ in three dimensions:
\begin{align}
&a_{3}=R_{\rm e}\exp[2\pi/3\tilde{f}(0)], & D=1,\label{a31D}\\
&S_3=-3\tilde{f}(0)/4\pi^2, & D=2,\label{a32D}\\
&\Upsilon_3=-3\tilde{f}(0)/4\pi^3, & D=3.\label{a33D}
\end{align} 
It is useful to note that for $D=2,3$ the three-body potential $g_3\delta(\sqrt{3}{\bf x}/2)\delta({\bf y})$ with \cite{RemCoeff}
\begin{equation}\label{g3throughf0}
g_3=-3(\sqrt{3}/2)^D\tilde{f}(0)
\end{equation} 
treated in the first Born approximation would produce the same scattered wave as Eqs.~(\ref{PsiAsymp}). Equations~(\ref{a32D}), (\ref{a33D}), and (\ref{g3throughf0}) relate the three-body coupling constant $g_3$ to the three-body scattering surface and hypervolume. The corresponding contribution to the energy density of a three-body-interacting condensate equals $g_3n^3/6$ in the weakly interacting regime, which is defined by $|S_3|n\ll 1$ in two dimensions and by $|\Upsilon_3|n^{4/3}\ll 1$ for $D=3$. The quantity $g_3/L^{2D}$ gives the energy shift for three (condensed) atoms in a large volume $L^D$. By solving the three-body problem nonperturbatively we calculate the exact $g_3$, which can then be compared to the mean-field result given by Eq.~(\ref{g3}).

The relation between $a_3$ and the three-body energy shift in the case $D=1$ is slightly more subtle. Pastukhov~\cite{Pastukhov2019} has recently shown that the ground-state energy density of a three-body-interacting one-dimensional Bose gas can be expanded in half-integer powers of the small parameter 
\begin{equation}\label{g3Vertex}
g_3(n)=\sqrt{3}\pi/\ln(1/a_3n)\ll 1,
\end{equation}
with the leading-order term equal to $E/L=g_3(n)n^3/6$. Although, $g_3$ given by Eq.~(\ref{g3Vertex}) depends on $n$, one can replace $1/n$ by another density-independent length scale $l$. If this scale is not exponentially different from $1/n$, the two small parameters are equivalent since they differ only by a higher-order term $\sim g_3^2$. By computing $a_3$ we can thus compare Eqs.~(\ref{g3}) and (\ref{g3Vertex}) which we expect to approach each other in the limit $R_{\rm e}/a_{12}\rightarrow 0$ (at fixed $n$). Equivalently, one can say that in this limit Eq.~(\ref{g3}) predicts the leading exponential dependence of the one-dimensional three-body scattering length
\begin{equation}\label{a3mf}
a_3\propto \exp\left(\frac{\pi}{\sqrt{3}}\frac{\mu_{12}a_{12}}{R_{\rm e}}\right)=\exp\left(\frac{2\pi}{3\sqrt{3}} \frac{a_{12}}{R_{\rm e}}\right)
\end{equation}
leaving, however, the preexponential factor unknown.

Returning to the task of determining $\tilde{f}(0)$ from Eqs.~(\ref{Integr}) and (\ref{Differ}) we note that the three-body problem in hand admits a zero-range description parametrized by $a_{11}$, $a_{12}$, and $R_{\rm e}$ (see, however, Sec.~\ref{ThreeBody3D}). Indeed, the sum $\hat{L}f(x)+f(x)/g_{11}$ in Eq.~(\ref{Integr}) is well behaved in the limit $r_0\rightarrow 0$ since the singularity of $\hat{L}f(x)$ gets canceled by the $r_0$-dependent term in $1/g_{11}$ [see Eqs.~(\ref{g2D}) and (\ref{g3D})]. The parameter $r_0$ thus drops out from Eq.~(\ref{Integr}), $g_{11}$ being conveniently eliminated in favor of $a_{11}$. As far as Eq.~(\ref{Differ}) is concerned, one can just substitute the interaction term $g_{12}\tilde\delta_{r_0}(\sqrt{3}x/2)$ by the Bethe-Peierls boundary conditions at $x\rightarrow 0$ 
\begin{align}
&\phi(x)\propto |x|-2a_{12}/\sqrt{3}, & D=1,\label{BP1D}\\
&\phi(x)\propto \ln(\sqrt{3}x/2a_{12}), & D=2,\label{BP2D}\\
&\phi(x)\propto 1-2a_{12}/\sqrt{3}x, & D=3.\label{BP3D}
\end{align}
In other words, Eq.~(\ref{Differ}) is equivalent to 
\begin{equation}\label{DifferZeroRange}
-\nabla^2_{\bf x}\phi({\bf x})=-\sqrt{2/R_{\rm e}^{D}}f({\bf x})
\end{equation} 
supplemented by the boundary conditions (\ref{BP1D})-(\ref{BP3D}).

From now on, for brevity, we choose to measure all distances in units of $R_{\rm e}$. The function $\tilde{f}(0)$ then depends on $a_{11}$ and $a_{12}$ (measured in units of $R_{\rm e}$) and its dimension is clear from Eq.~(\ref{PsiAsymp}).

The idea of solving Eqs.~(\ref{Integr}) and (\ref{BP1D})-(\ref{DifferZeroRange}) is to eliminate $\phi$ by inverting the Laplacian in Eq.~(\ref{DifferZeroRange}) and then deal with a single integral equation for $f$. We perform this procedure in momentum space [the Fourier transform is defined by $\tilde{F}(p)=\int_{\bf x} F(x)e^{-i{\bf px}}$] where Eq.~(\ref{DifferZeroRange}) formally transforms into $p^2\tilde{\phi}(p)=-\sqrt{2}\tilde{f}(p)$. Note, however, that we can always add to $\phi({\bf x})$ a general solution of the Laplace equation $-\nabla^2_{\bf x}\phi=0$, possibly singular at the origin. The solution of Eq.~(\ref{DifferZeroRange}) in momentum space is thus $-\sqrt{2}\tilde{f}(p)/p^2$ plus any linear combination of $\delta({\bf p})$ and $1/p^2$. The freedom of choosing the corresponding coefficients is removed by Eq.~(\ref{Integr}) and the boundary conditions (\ref{BP1D})-(\ref{BP3D}). The passage to momentum space in Eq.~(\ref{Integr}) is realized by rewriting the Fourier-space version of the operator
\begin{widetext}
\begin{equation}\label{LFourier}
(\hat{L}+1/g_{11})\tilde{f}(p)=\left(\frac{2}{\sqrt{3}}\right)^{D-2}\sum_\pm \int \frac{\tilde{f}(q)}{p^2\pm {\bf pq}+q^2}\frac{d^Dq}{(2\pi)^D}+\tilde{f}(p)\times\left\{\begin{aligned}&1/2|p|-a_{11}/2,\; &D=1,\\
&-(1/2 \pi) {\rm ln} (pa_{11}e^\gamma / 2),\; &D=2,\\
&-p/4\pi+1/4\pi a_{11},\; &D=3.
\end{aligned}
\right.
\end{equation}
\end{widetext}

We now proceed to reformulating the boundary conditions (\ref{BP1D})-(\ref{BP3D}) in momentum space. To this end let us first study the large-$x$ behavior of $\phi(x)$ and $f(x)$ and check that these functions indeed possess well-defined Fourier transforms. When two atoms are far away from the third one (large $x$), the function $\phi$ is approximately proportional to $\Psi$ due to Eq.~(\ref{TwoBodyphi}), which is equivalent to having small $f$ in Eq.~(\ref{Psi0throughf}). Thus, the large-$x$ asymptotic behavior of $\phi(x)$ is given by Eq.~(\ref{PsiAsymp}) and, by calculating the second derivative of these asymptotes and using Eq.~(\ref{DifferZeroRange}), we obtain the large-$x$ scaling $f(x)\propto x^{-2D}$. We conclude that the passage to momentum representation is straightforward for $D>1$ where $f(x)$ and $\phi(x)$ are well behaved. By contrast, in one dimension $\phi(x)\propto \ln|x|$ should be understood in the generalized sense by using a limit of a series of Fourier-transformable functions. In particular, we can use the relation $K_0(\sqrt{\epsilon}|x|)\approx -\ln\frac{\sqrt{\epsilon}|x|e^\gamma}{2}$ valid for small $\epsilon>0$ and define a generalized Fourier transform of $\ln|x|$ as 
\begin{equation}\label{Fourier}
-\frac{\pi}{|p|}=\lim_{\epsilon\rightarrow +0}\left[-\frac{\pi}{\sqrt{p^2+\epsilon}}-2\pi\delta(p)\ln\frac{\sqrt{\epsilon}e^\gamma}{2}\right].
\end{equation}
An immediate application of this formalism is the reformulation of the Bethe-Peierls boundary condition (\ref{BP1D}) in momentum space. Namely, for small $x$ we have
\begin{equation}\label{BP1DFourier}
\phi(x)= \int \tilde{\phi}(p)\frac{dp}{2\pi} - \frac{|x|}{2}\lim_{p\rightarrow \infty}p^2\tilde{\phi}(p)+o(x),
\end{equation}
where the integral is convergent, the singularity $\tilde{\phi}(p)\propto 1/|p|$ being understood in the sense of Eq.~(\ref{Fourier}). Comparing Eq.~(\ref{BP1DFourier}) with (\ref{BP1D}) and denoting $C=\lim_{p\rightarrow \infty}p^2\tilde{\phi}(p)$ gives us the Bethe-Peierls boundary condition in momentum space
\begin{equation}\label{BP1DFin}
\int \frac{\tilde{\phi}(p)}{C}\frac{dp}{2\pi}=\frac{a_{12}}{\sqrt{3}}.
\end{equation}
Repeating the same procedure in two dimensions Eq.~(\ref{BP2D}) transforms into
\begin{equation}\label{BP2DFin}
\int \left[\frac{\tilde{\phi}(p)}{C}-\frac{1}{p^2+\sigma}\right]\frac{d^2p}{(2\pi)^2}=\frac{1}{2\pi}\ln\frac{a_{12}\sqrt{\sigma}e^\gamma}{\sqrt{3}},
\end{equation}
where $\sigma$ is any positive number. In the case $D=3$, Eq.~(\ref{BP3D}) becomes
\begin{equation}\label{BP3DFin}
\int \left[\frac{\tilde{\phi}(p)}{C}-\frac{1}{p^2}\right]\frac{d^3p}{(2\pi)^3}=-\frac{\sqrt{3}}{8\pi a_{12}}.
\end{equation}
The task of reformulating our problem in momentum space is thus over.

We now write the solution of Eq.~(\ref{DifferZeroRange}) in the form
\begin{equation}\label{DifferSol}
\tilde{\phi}(p)=\sqrt{2}(2\pi)^D\delta({\bf p})+\frac{C-\sqrt{2}\tilde{f}(p)}{p^2}.
\end{equation}
Equation~(\ref{DifferSol}) is consistent with the definition of $C$ (which is still unknown) and the coefficient in front of $\delta({\bf p})$ is dictated by Eq.~(\ref{Integr}) and by the fact that the operator (\ref{LFourier}) does not give rise to a delta function. We now eliminate $\tilde{\phi}(p)$ by substituting Eq.~(\ref{DifferSol}) into Eqs.~(\ref{Integr}) and (\ref{BP1DFin})-(\ref{BP3DFin}) and after simple manipulations we obtain the following results. 

\subsection{One dimension}

In one dimension we arrive at
\begin{equation}\label{f01D}
\tilde{f}(0)=\frac{1}{a_{12}/\sqrt{3}+I^{(1)}(a_{11})},
\end{equation}
where the function $I^{(1)}(a_{11})=\int \frac{\chi(p)-1}{p^2}\frac{dp}{2\pi}$ is defined through the solution of
\begin{equation}\label{chi1D}
\frac{\sqrt{3}}{2}\sum_\pm \int \frac{\chi(q)}{p^2\pm pq+q^2}\frac{dq}{2\pi}+\left(\frac{1}{2|p|}+\frac{1}{p^2}-\frac{a_{11}}{2}\right)\chi(p)=\frac{1}{p^2},
\end{equation}
$C=\sqrt{2}\tilde{f}(0)$, and $\tilde{f}(p)=\tilde{f}(0)\chi(p)$. Substituting Eq.~(\ref{f01D}) into Eq.~(\ref{a31D}) $a_3$ factorizes into (we restore the dimensions here)
\begin{equation}\label{a31Dres}
a_3=R_{\rm e}\exp\left(\frac{2\pi}{3\sqrt{3}}\frac{a_{12}}{R_{\rm e}}\right)\exp\left[\frac{2\pi}{3}I^{(1)}\left(\frac{a_{11}}{R_{\rm e}}\right)\right],
\end{equation} 
consistent with Eq.~(\ref{a3mf}) in the limit of small $R_{\rm e}/a_{12}$.

\begin{center}
\begin{figure}[ht]
\vskip 0 pt \includegraphics[clip,width=1\columnwidth]{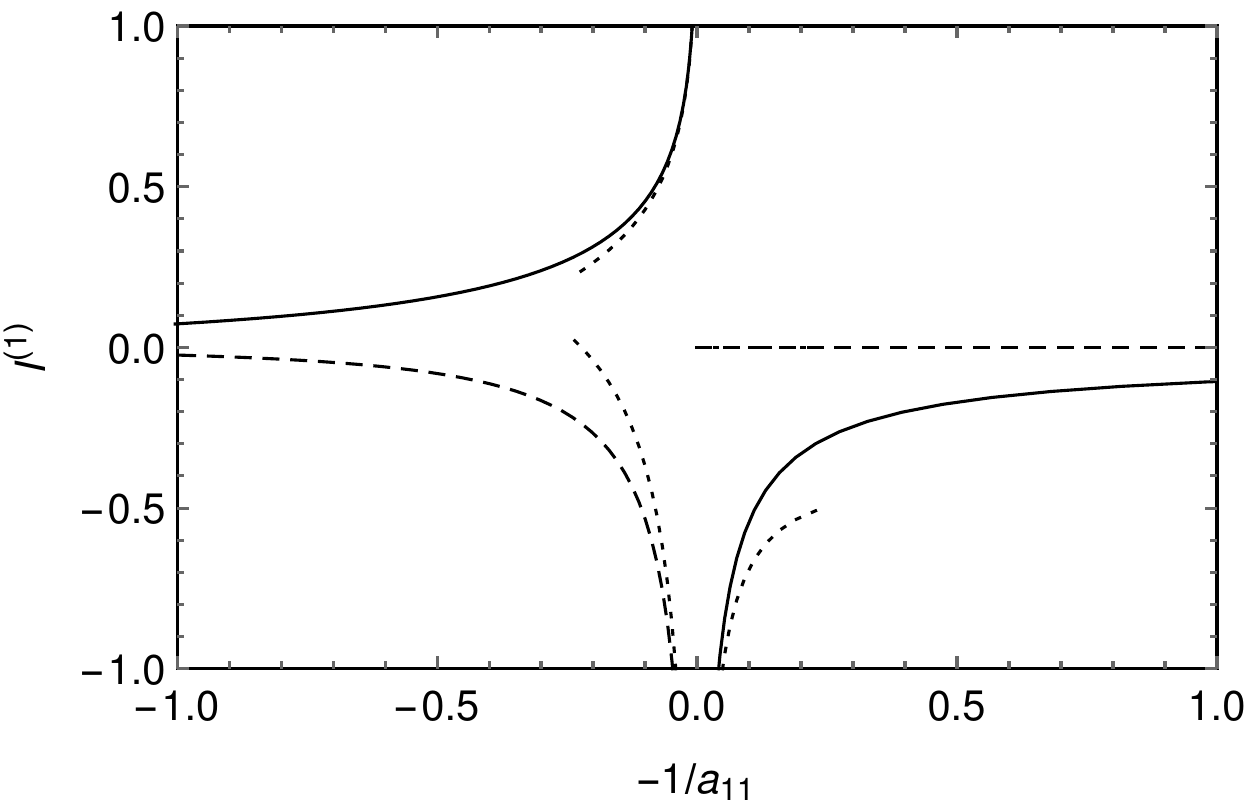}
\caption{
Functions ${\rm Re}I^{(1)}$ (solid) and ${\rm Im}I^{(1)}$ (dashed) characterizing the dependence of the effective three-body interaction on $a_{11}$ in one dimension [see Eqs.~(\ref{f01D}) and (\ref{PsiAsymp})]. $a_{11}$ is measured in units of $R_{\rm e}$. The dotted curves correspond to the large-$a_{11}$ asymptote [Eq.~(\ref{I1Dasymp})]. For $a_{11}\rightarrow \pm 0$ one has $I^{(1)} \approx -0.03$. 
}
\label{Fig:I1D}
\end{figure}
\par\end{center}

Let us now discuss the function $I^{(1)}$. For large $a_{11}$ (weak atom-atom interaction) this function can be expanded in powers of $\sqrt{-1/a_{11}}$. In order to see this we rescale the momentum $p=\sqrt{-1/a_{11}}z$ and rewrite Eq.~(\ref{chi1D}) in the form
\begin{align}
&\chi(z)=\frac{1}{1+z^2/2}\nonumber \\
&-\frac{1}{\sqrt{-a_{11}}}\frac{z^2}{1+z^2/2}\left[\frac{\sqrt{3}}{2}\sum_\pm \int \frac{\chi(y)}{z^2\pm yz+y^2}\frac{dy}{2\pi}+\frac{\chi(z)}{2|z|}\right],\label{chi1Drescaled}
\end{align}
which we then solve iteratively. In particular, the first iteration gives $\chi(z)= 1/(1+z^2/2)$ and provides the leading order term $I^{(1)}\approx -\sqrt{-a_{11}/8}$. The second iteration results in
\begin{equation}\label{I1Dasymp}
I^{(1)}= -\sqrt{-\frac{a_{11}}{8}}+\frac{9 + 5 \sqrt{3}\pi+27\ln (-a_{11}e^{-2\gamma}/2)}{36 \pi} +o(1).
\end{equation}
The solid and dashed lines in Fig.~\ref{Fig:I1D} show, respectively, the real and imaginary parts of $I^{(1)}$ as a function of $-1/a_{11}$ ($= g_{11}/2$) obtained numerically. The dotted lines indicate the real and imaginary parts of the large-$a_{11}$ asymptote (\ref{I1Dasymp}). 

For negative $a_{11}$ the solution is real and ${\rm Im}I^{(1)}\equiv 0$. By contrast, for $a_{11}>0$ the function $\chi(p)$ is characterized by simple poles at $p=\pm(\kappa+i0)$, where $\kappa>0$ is defined by Eq.~(\ref{kappa1D}) [this is also the point where the term in round brackets in Eq.~(\ref{chi1D}) vanishes]. These poles correspond to the three-body recombination to a dimer state, which, as found in Sec.~\ref{TwoBody}, exists only for positive $a_{11}$. One sees that $I^{(1)}$ and, therefore, $\tilde{f}(0)$ become complex reflecting the three-body loss. Technically, as one passes from positive to negative $-1/a_{11}$, the choice of the correct branch of the square root and logarithm in Eq.~(\ref{I1Dasymp}) is ensured by keeping $-1/a_{11}$ just below the real axis.

\subsection{Two dimensions}

The solution in the two-dimensional case can be written as
\begin{equation}\label{f02D}
\tilde{f}(0)=\frac{2\pi}{\ln(a_{12}e^\gamma /\sqrt{3})+2\pi I^{(2)}(a_{11})},
\end{equation}
where $I^{(2)}(a_{11})=\int \frac{\chi(p)-1/(p^2+1)}{p^2} \frac{d^2p}{(2\pi)^2}$ and $\chi$ satisfies
\begin{equation}\label{chi2D}
\sum_\pm \int \frac{\chi(q)}{p^2\pm {\bf pq}+q^2}\frac{d^2q}{(2\pi)^2}+\left(\frac{1}{p^2}-\frac{1}{2\pi}\ln\frac{a_{11}pe^\gamma}{2}\right)\chi(p)=\frac{1}{p^2}.
\end{equation}
The three-body scattering surface is proportional to $\tilde{f}(0)$ [see Eq.~(\ref{a32D})] and the mean-field result (\ref{g3}) is recovered for weak attractive or repulsive atom-dimer interactions (small or large $a_{12}$). As in the one-dimensional case we see that the dependence on $a_{12}$ is analytic and for the complete solution of the problem one needs to know only $I^{(2)}(a_{11})$. 

For a weak atom-atom background interaction (small or large $a_{11}$), introducing the small parameter $\lambda=1/\ln(1/a_{11})$, we can proceed iteratively in exactly the same manner as in the one-dimensional case. Namely, using the momentum rescaling $p=\sqrt{\lambda}z$ one can see that to the leading order $\chi(z)\approx 1/(1+z^2/2\pi)$ and after two iterations we have
\begin{equation}\label{I2Dasymp}
I^{(2)}= \frac{\ln(2\pi\lambda)}{4\pi}+\lambda\frac{\ln(C\lambda)}{8\pi}+o(\lambda),
\end{equation}
where $C\approx 0.013$.

\begin{center}
\begin{figure}[ht]
\vskip 0 pt \includegraphics[clip,width=1\columnwidth]{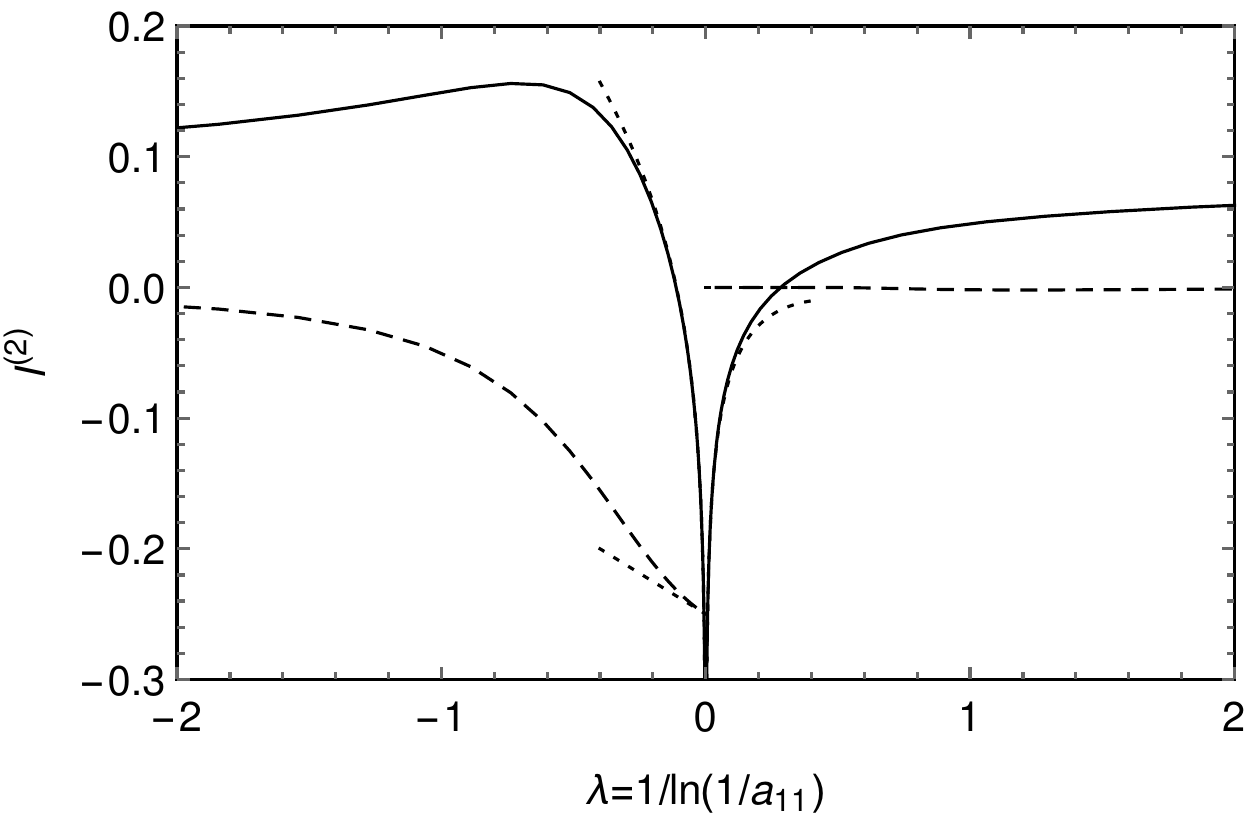}
\caption{
Real and imaginary parts of $I^{(2)}$ in the two-dimensional case. We use the same notations as in Fig.~\ref{Fig:I1D}. 
}
\label{Fig:I2D}
\end{figure}
\par\end{center}

In Fig.~\ref{Fig:I2D} we plot the real (solid) and imaginary (dashed) parts of $I^{(2)}$ versus $\lambda$ together with the asymptote (\ref{I2Dasymp}) (dotted). In the two-dimensional case ${\rm Im} I^{(2)}$ is always finite since there is always a dimer bound state available for the recombination (see Sec.~\ref{TwoBody}). However, for small positive $\lambda$ the dimer is exponentially deep and small (its energy is proportional to $1/a_{11}^2=e^{-1/\lambda}$) so that the recombination in this limit is not captured by the power expansion Eq.~(\ref{I2Dasymp}). 

Note that for small $\lambda$ the characteristic momentum involved in the solution $\chi(p)$ is $\sqrt{\lambda}$. Therefore, the asymptotic expansion (\ref{I2Dasymp}) is also valid if, instead of the zero-range atom-atom interaction, we have a potential of a finite but sufficiently small range $\ll 1/\sqrt{\lambda}=\sqrt{|\ln(1/a_{11})|}$, characterized by the same scattering length $a_{11}$. In particular, one can have a purely repulsive potential which does not lead to a dimer state in our problem.

\subsection{Three dimensions}\label{ThreeBody3D}

In three dimensions we have 
\begin{equation}\label{f03D}
\tilde{f}(0)=\frac{1}{-\sqrt{3}/8\pi a_{12}+I^{(3)}(a_{11})},
\end{equation}
where $I^{(3)}(a_{11})=\int \frac{\chi(p)}{p^2}\frac{d^3p}{(2\pi)^3}$ with $\chi$ satisfying
\begin{equation}\label{chi3D}
\frac{2}{\sqrt{3}}\sum_\pm \int \frac{\chi(q)}{p^2\pm {\bf pq}+q^2}\frac{d^3q}{(2\pi)^3}+\left(\frac{1}{p^2}-\frac{p}{4\pi}+\frac{1}{4\pi a_{11}}\right)\chi(p)=\frac{1}{p^2}.
\end{equation}
Here we also manage to separate the dependencies on the atom-dimer and atom-atom interactions. The mean-field solution (\ref{g3}) is retrieved for $a_{12}\rightarrow 0$. Calculating $I^{(3)}$ is, however, more subtle than in the low-dimensional cases. Indeed, small hyperradii effectively correspond to high collision momenta and energies where the two-body scattering length is approximated by its background value $a_{11}$. Thus, at $\rho\ll a_{11}$ we deal with the Efimovian three-boson system which requires a three-body parameter or a cutoff momentum. Mathematically, this can be seen from Eq.~(\ref{chi3D}) at momenta $p\gg 1/a_{11}$, where the dominant terms are the integral and $-p\chi(p)/4\pi$. The corresponding large-momentum behavior of $\chi(p)$ is a linear combination of Efimov waves $p^{\pm is_0-2}$ with $s_0\approx 1.00624$ \cite{Braaten2006}. The coefficients in this linear combination are fixed by introducing an external (three-body) parameter, phase, or momentum. Namely, one can set 
\begin{equation}\label{Efimov}
\chi\propto \frac{\sin[s_0\ln(p/p_0)]}{p^2}
\end{equation}
as the asymptotic boundary condition for $p\gg 1/a_{11}$. Accordingly, the quantity $I^{(3)}$ is, in fact, a function of $a_{11}$ and the three-body parameter $p_0$. However, for small $a_{11}$ the leading-order contribution to $I^{(3)}$ is universal, i.e., independent of $p_0$. Indeed, for small $a_{11}$ and momenta $p\ll 1/|a_{11}|$ Eq.~(\ref{chi3D}) reduces to $(1/p^2+1/4\pi a_{11})\chi(p)=1/p^2$. The corresponding solution $\chi= 1/(1+p^2/4\pi a_{11})$ is characterized by the typical momentum $\sqrt{a_{11}}\ll 1/a_{11}$ and leads to 
\begin{equation}\label{I3Dasymp}
I^{(3)}\approx \sqrt{a_{11}/4\pi}.
\end{equation} 
In order to estimate the next-order term we match $\chi(p)$ with the Efimov wave (\ref{Efimov}) at momentum $p\sim 1/|a_{11}|$ obtaining a contribution to $I^{(3)}$ of the order of $a_{11}^2$.

It makes sense to study the case of larger $a_{11}$ ($\gtrsim R_{\rm e}$) within our zero-range model, if we deal with a zero crossing near a narrow Feshbach resonance (large $R_{\rm e}$) which, in turn, lies in the vicinity of a broader Feshbach resonance (large $a_{11}$). At the same time it is interesting to have a significant atom-dimer interaction (large $a_{12}$) such that the two terms in the denominator of Eq.~(\ref{f03D}) are comparable. Then, in order to find the effective three-body force we also need to know the three-body and inelasticity parameters (or, equivalently, the real and imaginary parts of $p_0$), which could be known from the Efimov loss spectroscopy near the broad resonance. Given the large number of parameters in this problem we just give a prescription for calculating $I^{(3)}$. Namely, one has to solve Eq.~(\ref{chi3D}) with the boundary condition (\ref{Efimov}) at $p\rightarrow \infty$ also requiring $\chi\propto 1/(p-\kappa-i0)$ near the pole given by Eq.~(\ref{kappa3D}).   

\section{Discussion and conclusions}

In this article we have expanded the idea that the bosonic model with a Feshbach-type atom-dimer conversion (\ref{Ham}) near a two-body zero crossing can be reduced to a purely atomic model with an effective three-body interaction, which strongly depends on the atom-dimer conversion amplitude. As a particular example, we show that this mechanism of generating three-body forces can be used for stabilizing supersolid phases of two-dimensional dipoles. 

Sections~\ref{TwoBody} and \ref{ThreeBody} have been devoted to constructing a zero-range regularized version of the model (\ref{Ham}) with a minimal set of parameters ($a_{11}$, $a_{12}$, and $R_{\rm e}$). We have solved this model nonperturbatively in the two-body and three-body cases in all dimensions at the two-body zero crossing. Formulas (\ref{f01D}), (\ref{f02D}), and (\ref{f03D}) give analytic dependencies of the three-body scattering amplitude on $a_{12}$ in different dimensions. The dependence on $a_{11}$ is found numerically and also analytically for weak atom-atom background interactions. In the three-dimensional case, our three-body zero-range model is Efimovian and requires an additional three-body parameter. We find, however, that for small $|a_{11}|/R_{\rm e}$, effects associated with the Efimov physics are subleading.

These results show that for comparable and weak atom-dimer and atom-atom interactions (characterized by $g_{12}$ and $g_{11}$, respectively), the three-body interaction is mostly influenced by $g_{12}$, consistent with the mean-field result (\ref{g3}). However, the convergence is not always uniform. For example, in the two-dimensional case, one can simultaneously decrease $g_{12}$ and $g_{11}$, keeping both terms in the denominator of Eq.~(\ref{f02D}) comparable to (or even canceling) each other (resulting in a diverging three-body scattering surface). In the same spirit, we can use the nonperturbative three-dimensional formula Eq.~(\ref{f03D}) and predict a three-body resonance at $\sqrt{3}R_{\rm e}/8\pi a_{12}\approx \sqrt{a_{11}/4\pi R_{\rm e}}\ll 1$. 

Inelastic three-body events manifest themselves through the appearance of an imaginary part of $\tilde{f}(0)$, which, in turn, comes from the complex $I^{(D)}$ or complex atom-dimer scattering length $a_{12}$. The former reflects the three-body recombination to a dimer state and the latter the relaxation process in collisions of atoms with closed-channel dimers.

Several proposals on how to observe elastic three-body interactions experimentally are based on the following ideas. A repulsive three-body force could stabilize a system with attractive two-body interactions and make it self-trapped \cite{Bulgac2002}. The structure and energies of few-body bound states, detectable spectroscopically, are also influenced by these forces \cite{Sekino2018,Nishida2018,Pricoupenko2018,Guijarro2018}. Collective-mode frequency shifts in a trapped gas could be another experimentally observable signature of three-body interactions \cite{ValientePastukhov2019}.

\section*{Acknowledgements} The research leading to these results received funding from the European Research Council (FP7/2007--2013 Grant Agreement No. 341197) and we acknowledge support from ANR grant Droplets No. ANR-19-CE30-0003-02.


\begin{thebibliography}{99}
\bibitem{RemCoeff}
The prefactor $\sqrt{3}/2$ here [and also in Eqs.~(\ref{Differ}), (\ref{BP1D}-\ref{BP3D})] comes from our definition of the Jacobi coordinates (\ref{Jacobi}), according to which the atom-dimer distance equals $\sqrt{3}x/2$.
\end{thebibliography}
\end{document}